# Tunable Valley Polarization in Diamond

Nattakarn Suntornwipat[1*], Jan Isberg[1] and Saman Majdi[1]

[1]Department of Electrical Engineering, Uppsala University, Box 65, SE-751 03, Uppsala, Sweden

**Abstract**

Device stability is essential for quantum information technologies, where reliable control of electronic states is crucial. Diamond valleytronics offers a promising platform by exploiting the valley degree of freedom to store and manipulate information. In this work, we demonstrate a diamond-based valley transistor with a dual-gate, two-drain architecture that enables tunable valley-polarized transport via gate voltage modulation. By leveraging the significant effective-mass anisotropy of diamond's conduction band valleys, this architecture provides control over spatial distribution and transit times. We further demonstrate that valley-polarized transport in diamond is remarkably robust against thermal variations over macroscopic distances. These results demonstrate the resilience of valley states and highlight diamond's potential for energy-efficient valleytronic devices in next-generation quantum and high-power electronics.

KEYWORDS: Valleytronics, valley-polarized, diamond, field effect transistor, quantum transport

*Nattakarn Suntornwipat (Nattakarn.suntornwipat@angstrom.uu.se)

## Introduction

The valley degree of freedom, or *valley pseudospin*, has emerged as a novel quantum number for encoding and manipulating information in electronic systems. In multivalley semiconductors, electrons occupy distinct momentum-space valleys that can be selectively addressed using optical, electrical, or mechanical means. This concept is fundamental to *valleytronics,* which seeks to exploit valley polarization as an information carrier, analogous to spin in spintronics, offering pathways to ultrafast logic gate operations [1] and integration with quantum technologies [2].

Two-dimensional (2D) materials such as transition metal dichalcogenides (TMDs) and graphene have been widely studied, where broken inversion symmetry and strong spin–orbit coupling enable spin-valley locking and valley-selective optical excitation [3–7]. These properties have enabled demonstrations of valley-based transistors, logic gates, and optoelectronic devices [1,8,9]. However, valleytronics is not confined to 2D systems. Three-dimensional (3D) materials, including silicon, germanium, and diamond, also exhibit multiple conduction band valleys that can be polarized under strain, electric fields, or temperature gradients [10–12]. These findings highlight the potential of bulk materials, which offer complementary advantages to atomically thin systems, including enhanced thermal stability, mechanical robustness, and compatibility with established semiconductor processing.

Among these, diamond is particularly compelling due to its wide bandgap, exceptional thermal conductivity, and high breakdown voltage, make it suitable for operation in extreme environments. Diamond hosts six equivalent conduction band valleys along the <100> directions, which can be polarized through external electric fields or via ultrafast laser excitation, enabling valley-controlled transport [11,13,14]. Recent studies demonstrate that linearly polarized femtosecond infrared pulses can generate an oscillating electric field that accelerates electrons, inducing intervalley scattering and allowing for room-temperature valley manipulation without relying on strain or temperature gradients [14]. Furthermore, the long valley relaxation times demonstrated in diamond at low temperatures [15] provide opportunities for energy-efficient and potentially hybrid quantum-classical devices.

## Methods

Experimental demonstrations of diamond-based valley field-effect transistors (FETs) have shown that valley currents can be separated using dual-gate architectures [13]. In these devices, electrons are photo-generated, and their transport paths and transit times are influenced by gate voltages, enabling the detection of distinct valley-polarized populations at different contacts. The exceptional stability of valley states in diamond, attributed to its strong carbon–carbon bonding and the low defect density of high-quality crystals, supports long valley lifetimes required for such transport. While these advances established a foundation for exploring valley-dependent phenomena, prior research in Ref. [13] has primarily focused on separation of valley ensembles at specific operating points. To date, the use of such devices for probing the systematic modulation and thermal resilience of valley-polarized transport remains largely unexplored.

In this work, we extend prior efforts by systematically analyzing valley-polarized transport in diamond under varying gate voltages and temperatures. Our measurements reveal the interplay between drift, diffusion, and valley relaxation, providing a clear picture of the mechanisms that control valley selectivity. Building on recent advances in valley control in diamond, this study establishes key design principles for robust, energy-efficient valleytronic devices.

The devices were fabricated on freestanding single-crystal chemical vapor deposition (CVD) diamond plates (4.5 × 4.5 mm, thickness 490 and 510 μm). These samples possess ultralow nitrogen impurity concentrations (<0.05 ppb) to minimize ionized impurity scattering and ensure high carrier mobility required for valley-polarized transport. Prior to device fabrication, samples were cleaned in a mixed acid solution ($HNO_3$:$H_2SO_4$:$HClO_4$) for 40 minutes at 180 °C and followed by oxygen-terminated via low-power oxygen plasma for 60 seconds. A 30 nm $Al_2O_3$ layer was then deposited by atomic layer deposition (ALD) at 300 °C using trimethylaluminum and ozone as precursors in a Picosun R200 system. This serves as both gate dielectric and a surface passivation layer to reduce interface trap density and surface scattering. Contacts were patterned via photolithography and etched with hydrofluoric acid, followed by Ti/Al (20 nm / 300 nm) metallization for the source, gate, and drain electrodes. Finally, a semi-transparent Au layer (10 nm) was deposited on the back surface to provide a uniform back contact. The final device architecture, comprise a source electrode, two independently biased gate electrodes, and two drain electrodes. To clarify the layout of the dual-gate, two-drain structure, Figure 1 provides a schematic of the completed device.

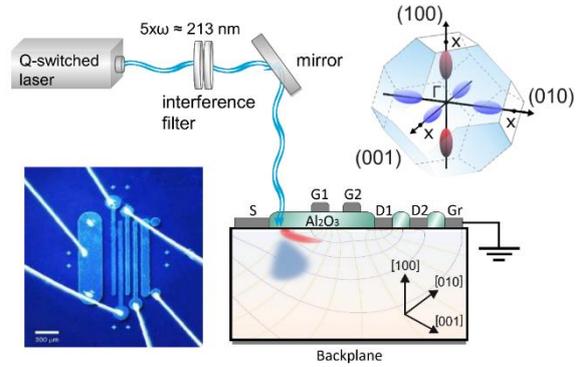

Figure 1: Schematic of the dual-gate, two-drain device illustrating electrostatic control of valley-polarized packets. Carriers generated by a 213 nm laser induce instantaneous currents in the drains as they traverse the channel, in accordance with the Shockley–Ramo theorem. The device includes an $Al_2O_3$ dielectric and a backplane electrode to modulate transport along the [100], [010], and [001] axes. (Top right) The inset displays the six equivalent conduction band valleys in the diamond Brillouin zone. (Bottom left) Image of the device from a top-down view showing the electrode geometry.

The devices were subsequently placed in a custom Janis ST-300MS vacuum cryostat with optical access. Carrier generation was performed near the source edge using a passively Q-switched DPSS laser emitting 213 nm pulses (FWHM: 800 ps, repetition rate: 313 Hz), with a photon energy of 5.82 eV to generate electron–hole pairs. To avoid excess carrier–carrier scattering and sample heating, the laser output was attenuated to below 1 nJ per pulse, maintaining a carrier density below $10^{10}$ cm$^{-3}$ and a thermal load under 0.3 μW. Under these conditions, electrons initially occupy all six conduction band valleys uniformly. Applying negative bias to the source, gates, and back contact drives valley-selective electron drift toward the drain electrodes, while holes are rapidly collected at the source, rendering their effect negligible. In accordance with the Shockley–Ramo theorem, the moving electrons induce instantaneous currents in the drain electrodes as they traverse the sample. Drain currents were recorded at virtual ground to ensure measurement stability. Temperature stability within ±0.1 K was maintained using a LakeShore 331 controller and a calibrated GaAlAs diode sensor (TG-120-CU-HT-1.4H) in good thermal contact with the sample. Current signals were amplified by broadband low-noise amplifiers (1GHz) and captured with a digital sampling oscilloscope (3 GHz bandwidth, 10 GS/s sampling rate).

## Results and discussion

The conduction band of diamond, much like that of silicon, contains six equivalent conduction band valleys along the <100> directions in momentum space. The strong covalent bonding and exceptionally high phonon energy at the K-point result in a long intervalley scattering time, approximately 300 ns at 77 K [15]. This stability is governed by the $f$-type (between orthogonal valleys) and $g$-type (between opposite valleys) intervalley scattering rates, expressed as:[16]

$$\frac{1}{\tau_{f,g}} \propto \frac{D_{f,g}^2 (m^*)^{3/2}}{\rho \omega_{f,g}} \left[ N_{f,g} \sqrt{E + \hbar\omega_{f,g}} + (N_{f,g} + 1)\sqrt{E - \hbar\omega_{f,g}} \right] \quad (1)$$

Here $D_{f,g}$ is the intervalley deformation potential, $m^*$ is the density-of-states effective mass, $\rho$ is the mass density of diamond, $\omega_{f,g}$ is the angular frequency of the specific phonon mode. $N_{f,g}$ denotes the phonon occupation number, following the Bose-Einstein distribution. The terms $\sqrt{E + \hbar\omega_{f,g}}$ and $\sqrt{E - \hbar\omega_{f,g}}$ correspond to the density of final states available for phonon absorption and emission processes, respectively. Here $E$ is the initial carrier energy and $\hbar$ is the reduced Planck's constant.

In diamond, the phonons required for intervalley scattering have distinct energies: $\hbar\omega_f$ = 110 meV for scattering between orthogonal valleys and $\hbar\omega_g$ = 165 meV for scattering between opposite valleys. Since both energies are significantly higher than the thermal energy ($k_B T \approx 6.6$ meV at 77 K), the phonon occupancy is negligible, effectively freezing out intervalley transitions. This ensures that the valley polarization is robustly preserved during transport, provided the transit time is much shorter than the intervalley relaxation time.

In contrast, intravalley acoustic scattering, which dictates the mobility, occurs on a much shorter timescale, about 1 ps at 77 K [11]. The intravalley acoustic scattering rate, $1/\tau_{ac}$, is given by:

$$\frac{1}{\tau_{ac}} \approx \frac{\sqrt{2}(m^*)^{3/2} D_A^2 k_B T}{\pi \hbar^4 \rho v^2} \sqrt{E} \quad (2)$$

In this expression, $1/\tau_{ac}$ is the intravalley acoustic phonon scattering rate, which serves as the primary limitation on carrier mobility at cryogenic temperatures. $D_A$ is the acoustic deformation potential (approximately 12.0 eV for diamond), and $k_B$ is the Boltzmann constant. $v$ denotes the averaged sound velocity ($v \approx 1.8 \times 10^4$ m/s). The factor $\sqrt{E}$ represents the density of final states available for a carrier with energy. Notably, because $v$ is exceptionally high in diamond, the energy exchange during these collisions is significant relative to $k_B T$, requiring the treatment of these interactions as strongly inelastic rather than following the traditional elastic approximation [17]. Because diamond possesses the highest sound velocity and Debye temperature of any bulk semiconductor, these intravalley interactions remain strongly inelastic. This inelasticity results in the observed deviation from the conventional $T^{-3/2}$ mobility law [17].

This disparity in timescales allows electrons to occupy six relatively stable valley pseudospin states. Electrons confined within a single valley exhibit pronounced effective mass anisotropy, with the longitudinal mass ($m_l \approx 1.56 m_0$) roughly 5.5 times greater than the transverse mass ($m_t \approx 0.28 m_0$) [18]. This anisotropy enables the separation of electrons with distinct valley pseudospins under an appropriately applied electric field, forming the basis for the electrostatic control demonstrated in our dual-gate architecture.

To investigate the dynamics of valley-polarized charge transport, experiments are performed on a two-gate, two-drain diamond transistor (Figure 1) under varying electrical and thermal conditions to assess sample dependence and reproducibility. This design provides enhanced control over valley currents compared to single-gate structures, while multiple drain contacts enable spatially resolved detection of valley-polarized carriers. The specific contact configuration is described in detail in our previous work [13]. Measurements are conducted under three categories: voltage-dependence modulation of valley transport, assessment of sample dependence and reproducibility, and the impact of thermal variations on valley resilience.

The applied gate voltages play a critical role in controlling valley-polarized charge transport in diamond-based transistors. By changing the voltages at the contacts, valley-dependent transport is modulated, since valleys exhibit anisotropic effective masses, causing electrons in different valleys to accelerate differently under the same electric field. These influences both the arrival time and current density from each valley. The dual-gate configuration allows control of both lateral and vertical electric fields. Varying the voltages applied to different contacts provides insight into the coupling between valley states and the device geometry, as well as the stability of valley polarization under different bias conditions.

The total electrostatic potential within the device is determined by the linear superposition of potentials from each contact:

$\phi(\vec{r}) = \phi_{\text{source}}(\vec{r}) + \phi_{\text{leftgate}}(\vec{r}) + \phi_{\text{rightgate}}(\vec{r}) + \phi_{\text{backplane}}(\vec{r})$ (3)

where $\phi_{\text{source}}(\vec{r})$, $\phi_{\text{leftgate}}(\vec{r})$, $\phi_{\text{rightgate}}(\vec{r})$, $\phi_{\text{backplane}}(\vec{r})$ represent the potential due to the source, left gate voltage, right gate voltage, and backplane, respectively and $\vec{r} = (x, y, z)$. In our case, the electric field components are: $E_x = -\frac{\partial \phi}{\partial x}, E_y = -\frac{\partial \phi}{\partial y}$, where $x$ is in the lateral direction and $y$ is in the vertical direction.

To analyze the influence of gate voltages on carrier dynamics, the source voltage is fixed at 7 V, the back gate at 2 V, the right gate voltage varies between 2 and 4 V and the left gate voltage is swept from 0 to 14 V at 77 K. Under these conditions, two distinct carrier types are observed, consistent with previous reports [13]: the electrons in the (001) valleys confined near the surface and the electrons in the (010) and (100) valleys penetrating deeper into the bulk (Figure 1). This spatial partitioning is governed by the diffusion length, $L = \sqrt{D\tau}$, where the diffusion coefficient D is related to the effective mass via the Einstein relation. The heavy vertical effective mass ($m_l$) of the (001) valley suppresses vertical diffusion, effectively confining these carriers to a high-mobility surface channel. Conversely, the lighter vertical mass ($m_t$) and distinct relaxation times of the (100) and (010) valleys permit deeper penetration into the diamond bulk. This anisotropy establishes a dual-channel system: a high-mobility lateral population in the (001) valley and a more dispersed, lower-mobility population in the (100)/(010) valleys.

At low left-gate voltages (Figure 2), transport is primarily dominated by electrons in the (100) and (010) valleys. These carriers exhibit a deeper spatial distribution within the sample due to their light vertical effective mass and the relatively weak transverse electrostatic confinement. As the left-gate voltage increases, the resulting transverse field pushes these carriers further into the bulk. Consequently, the (001) valley, which has a heavy vertical mass, emerges at intermediate voltages at Drain 1 and higher voltages at Drain 2. Meanwhile, the (100) and (010) populations, initially detected at low bias, gradually shift toward Drain 2 as the increased bias modulates their trajectory. At sufficiently high voltages, these slower carriers either extend beyond the designed contact area r are collected at the backplane. The right gate voltage exhibits a distinct influence on transport behavior, primarily modulating the carriers as they approach the collection region. At approximately 2 V (Figure 2), both carrier types are observed at Drain 1. However, as the right gate voltage increases, the (001) valley intensity at Drain 1 decreases as these carriers are progressively shifted toward Drain 2. This contrast in behavior highlights the functional separation of the gates: while the left gate governs penetration depth at the point of injection, the right gate modulates the final spatial distribution between the drains. Notably, the right gate does not influence the initial threshold for the (001) valley signature, where we still observe the peak at a left-gate bias of approximately 5 V; this confirms that the early stages of transport are independent of the drain-side gate settings. Instead, the right gate acts as a final steering mechanism, shifting peak positions and intensities as the bias increases. We select a source bias of 7 V for these measurements to ensure sufficient velocity differential and minimal diffusive broadening, allowing for the clear temporal resolution of the distinct valley populations.

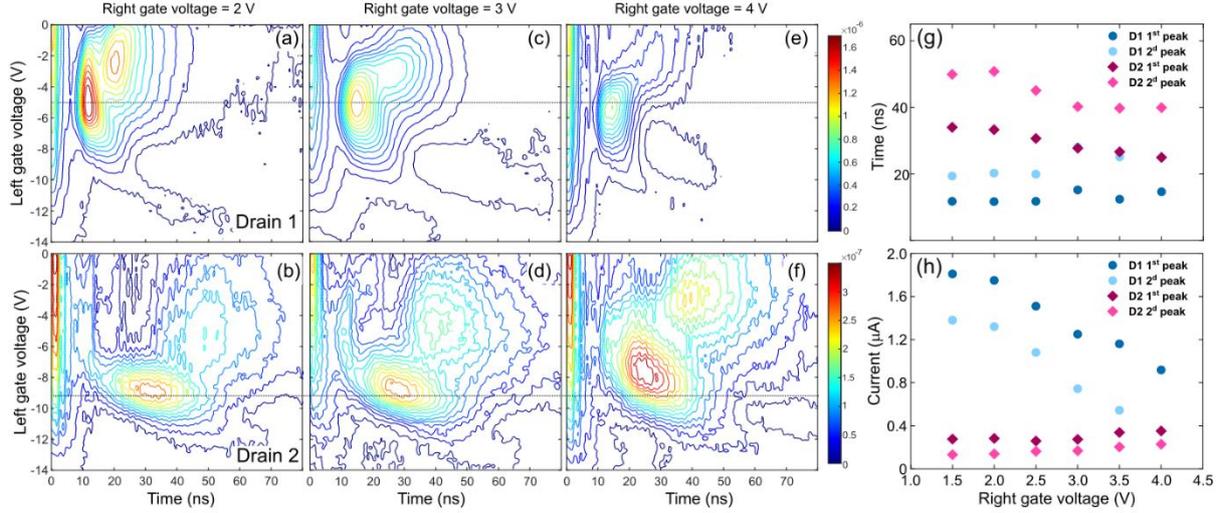

Figure 2: (a–f) Time-resolved current contour maps as a function of left-gate voltage for Drain 1 (top) and Drain 2 (bottom) at fixed right-gate voltage values of 2, 3, and 4 V. Color scale indicates induced current (µA). Horizontal dashed lines at left-gate voltages are included as visual guides to highlight the shift in the position of the first valley-polarized current peak as the right-gate voltage is varied. (g) Peak arrival times and (h) current amplitudes for the two observed valley populations at Drain 1 and Drain 2, plotted as a function of right-gate voltage at 77 K (source voltage: 7 V; backplane voltage: 2 V).

To further understand the role of the backplane, the source and right gate voltages were fixed at 7 and 3 V, respectively, while backplane voltage varies between 2 and 3 V. The left gate voltage is again swept from 0 to 14 V at 77 K to map the resulting three-dimensional electrostatic influence on valley selectivity.

Increasing the backplane voltage shifts the carrier distribution toward the top surface, counteracting the electrostatic confinement imposed by the top gates. A higher backplane bias increases the transverse electric field $E_y$, enhancing upward band bending and confining carriers near the surface. This reduces the penetration depth and decreases the induced current. At Drain 1 (Figure 3), the signal intensity drops, likely due to enhanced surface-roughness scattering or carrier loss near the interface. Drain 2 exhibits changes in both arrival timing and curve shape, reflecting the altered vertical distribution of the carriers.

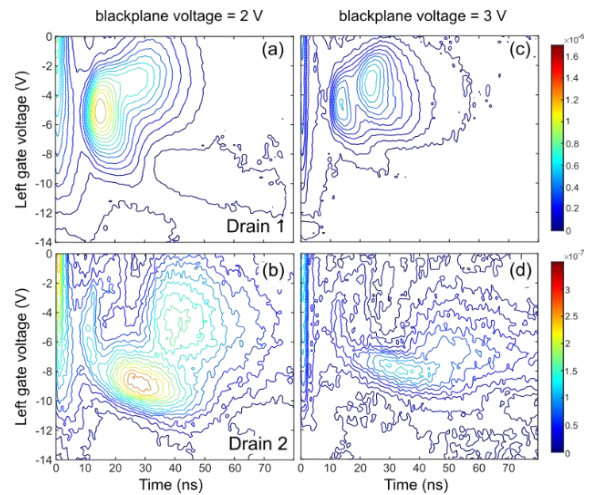

Figure 3: Impact of backplane modulation on carrier confinement. Time-resolved current contour maps acquired at 77 K for (a, c) Drain 1 and (b, d) Drain 2, showing the effect of increasing backplane voltage from 2 V to 3 V. Color scale indicates induced current (µA).

When both backplane and right gate voltages are varied simultaneously, the carrier transport behavior becomes more complex. Competing electric fields modify $E_x$ and $E_y$, limiting penetration depth and altering induced currents at both drains. This behavior can be explained using electrostatic modeling based on the Poisson equation:

$$\nabla^2 \Phi = \frac{q}{\varepsilon_0 \varepsilon_r}(\sum n_i - p) \quad (4)$$

where $n_i$ represents electron concentration in valley number 1, 2 or 3, p is hole concentration, $\varepsilon_r$ is the relative permittivity of diamond and $\varepsilon_0$ is the permittivity of vacuum.

The interplay between the backplane potential, which pushes carriers toward the surface, and the gate field, which drives them deeper into the bulk, dynamically reshapes the valley-dependent diffusion profiles. Additional measurement data for these simultaneous voltage sweeps are provided in the Supporting Information.

To evaluate the reproducibility and sample-dependent nature of the observed valley transport, measurements are performed on two distinct diamond samples under identical experimental conditions: source voltage of 7 V, back gate of 3.4 V, right gate of 3.4 V, and left gate is varied from 0 to 14 V at 77 K. Additionally, the same sample is measured under these identical conditions at different times to assess the temporal stability of the system.

The overall trends in the induced-current response remain consistent across different samples, though variations in peak intensity and voltage positions are observed. These differences are attributed to variations in intrinsic sample quality and interface characteristics, as evidenced by the time-of-flight (ToF) measurements shown in Figure 4(c). To establish a baseline, we refer to previous characterizations of these specific samples in a vertical configuration, where carriers traverse the bulk diamond without the influence of an oxide layer. Those measurements yielded room-temperature electron mobilities of 2190 cm²/V·s for Sample #1 and 2140 cm²/V·s for Sample #2, indicating a marginal bulk performance gap of approximately 2.3%.

However, in the lateral dual-gate architecture used in this study, transport dynamics are further influenced by the diamond/dielectric interface. As shown in Figure 4(a) and Figure 4(b), Sample #1 requires a higher left-gate voltage to reach its peak valley current compared to Sample #2. This shift suggests that Sample #1 possesses a higher density of interface traps or fixed charges within the $Al_2O_3$ layer. In metal-oxide-semiconductor field-effect transistors (MOSFETs), the interface trap density likely residing at the diamond-oxide interface below the gates. This difference in trap density between the two samples may stem from variations in surface roughness or morphology prior to the oxide deposition process. These traps effectively screen the applied gate field, requiring a higher bias to achieve equivalent carrier modulation. Despite these variations in interface trapping and threshold voltages, the fundamental valley-steering behavior and the relative velocity differential remain robust across different diamond samples.

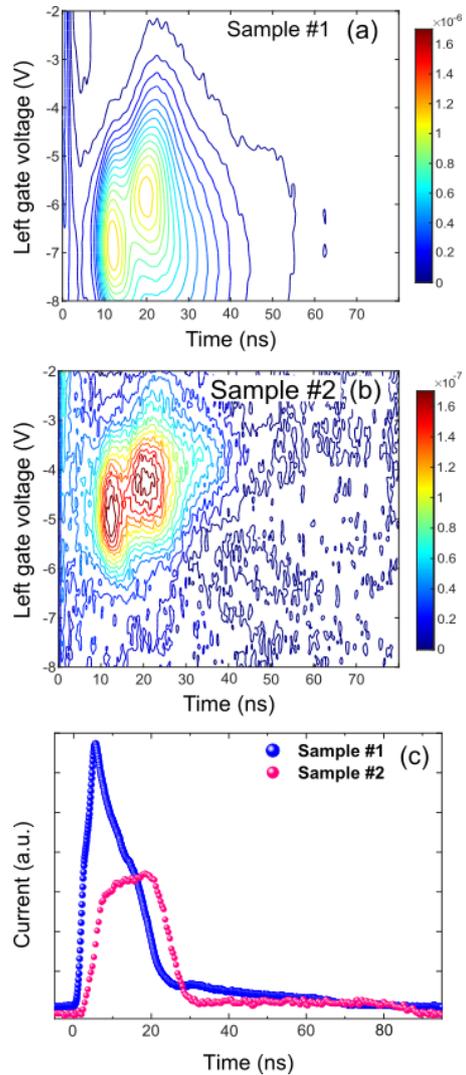

*Figure 4: Comparative time-resolved current contour maps for (a) Sample #1 and (b) Sample #2 under identical experimental conditions. (c) Time-of-flight measurements were performed on both samples at 77 K. The color bar represents the induced current (µA).*

To confirm the reliability of the experimental setup, additional measurements are performed on the same sample at one week apart. As shown in the Supporting Information in Figure S2, the signal profiles, shapes, and peak positions remain largely unchanged across these temporal trials. This confirms that the observed valley-polarized transport is a stable physical phenomenon and is not an artifact of experimental drift or environmental fluctuations.

Finally, we investigate the influence of the temperature on carrier transport by varying the temperature from 10 to 77 K under constant bias. At the lowest temperatures (Figure 5), carriers exhibit significantly shorter drift times, consistent with suppression of acoustic phonon scattering. As the temperature increases, the acoustic phonon population rises, leading to more frequent scattering events and a gradual increase in transit time. Quantitatively, we observe the peak arrival time shift from 8.7 ns at

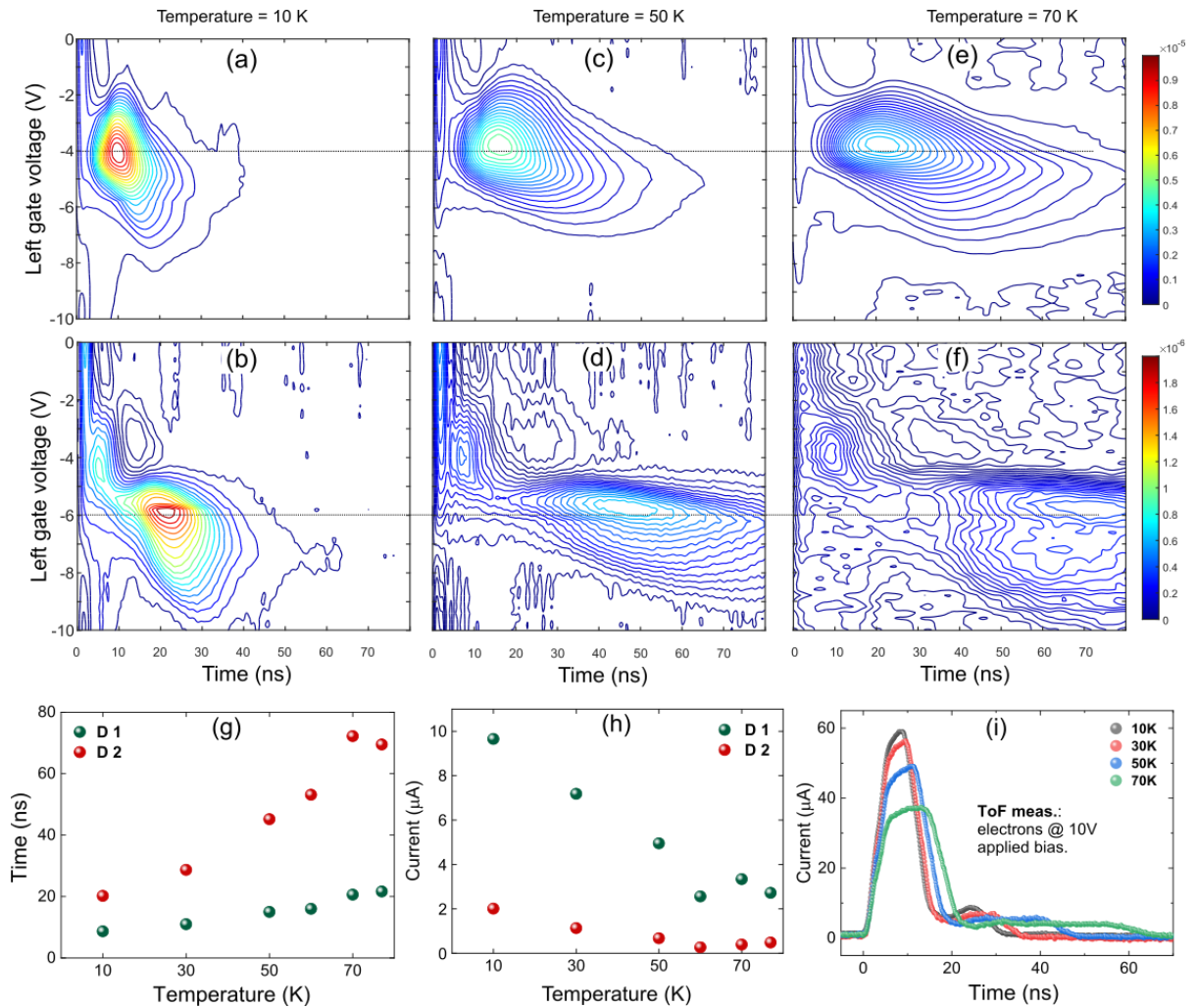

*Figure 5: Temperature dependence and thermal resilience of the valley populations. (a–f) Time-resolved current contour maps acquired at 10, 50, and 70 K for Drain 1 (top) and Drain 2 (bottom). Color scale indicates induced current (μA). Horizontal dashed lines at left-gate voltages are included as visual guides to highlight the shift in the position of the first valley-polarized current peak as the right-gate voltage is varied. (g) peak arrival times and (h) corresponding current amplitudes extracted from Drain 1 and Drain 2 over the temperature range of 10–77 K. (i) Current traces from ToF measurements were obtained on the same device from 10 to 70 K under an applied voltage of 10 V.*

10 K to 21.6 ns at 77 K, representing an increase by a factor of approximately 2.5.

These findings are consistent with the trends reported in Ref. [17], where a vertical ToF architecture is utilized with a travel distance of 510 μm and an electric field of 135 V/cm. In that study, drift times for fast and slow electrons are observed to shift from 12 to 18 ns (a factor of 1.5) and 28 to 62 ns (a factor of 2.2), respectively, over a similar temperature range. While our absolute drift times are shorter as expected given our reduced travel distance (300 μm) and the higher average electric field (around 187 V/cm for $V_s$ = 5.6 V), the relative temperature sensitivity remains comparable.

At the lower source bias of 5.6 V used for temperature-dependent studies, the induced current appears as a single, merged peak centroid. This merging is attributed to the reduced lateral electric field, which decreases the velocity differential between the (001) and (100)/(010) valleys, and the increased transit time, which enhances diffusive broadening. Consequently, the measured arrival time factor of 2.5 represents an ensemble-averaged response of the mixed valley populations. This factor, being slightly higher than the bulk slow carrier factor (2.2), likely reflects additional scattering sensitivity from the diamond/dielectric interface.

In addition to changes in drift time, the signal shape evolves with temperature. At 10 K, the induced current peak is relatively sharp and larger in amplitude, indicating minimal diffusion. As temperature rises to 77 K, the peak broadens, reflecting enhanced carrier diffusion due to increased thermal energy, and its amplitude decreases slightly. This diffusion-driven spreading causes carriers to arrive over a wider time window. Importantly, intervalley scattering remains minimal across this range, preserving valley polarization. However, at higher temperatures, the onset of additional features in the induced current suggests increased intervalley coupling, which could impact valley lifetime and device performance. These results underscore the advantage of operating diamond-based valley transistors at low temperatures for applications requiring long valley coherence times.

In conclusion, we have investigated valley-polarized charge transport in diamond-based dual-gate, two-drain transistors under controlled electrostatic and thermal conditions. By systematically varying gate voltages and temperature, we demonstrate tunable valley-polarized transport and spatial steering of valley populations, enabled by the pronounced effective-mass anisotropy of diamond's conduction-band valleys. The dual-gate architecture provides independent control of carrier penetration depth and lateral steering, allowing clear temporal and spatial discrimination of valley-polarized electron ensembles.

## Conclusion

Temperature-dependent measurements reveal that valley-polarized transport in diamond is remarkably robust against thermal variations. Carrier drift times increase by only a factor of approximately 2.5 between 10 and 77 K, consistent with the suppression of intervalley scattering due to diamond's high phonon energies and rigid lattice. Despite increased acoustic phonon scattering and diffusive broadening at elevated temperatures, valley polarization remains largely preserved over the relevant transport timescales. Measurements across multiple samples further confirm the reproducibility and stability of the observed transport behavior, indicating that valley-polarized transport is resilient to variations in sample quality and interface conditions.

These results establish diamond as a uniquely stable platform for solid-state valleytronics. The combination of electrostatic tunability and thermal robustness highlights the potential of diamond valley devices for energy-efficient electronic and hybrid quantum technologies. Beyond demonstrating controllable valley transport, this work provides a framework for designing valleytronic architectures in wide-bandgap semiconductors.

## Acknowledgements

This study is supported by ÅForsk Foundation (Grant number ÅF 19-427 and ÅF 21-53), the STandUP for Energy strategic research framework and the Swedish Research Council (Grant No. 04186-5). The device processing was performed at the Ångström Microstructure Laboratory (MSL), Uppsala University, a node of the Myfab national research infrastructure.